\documentclass[times, twocolumn,final]{elsarticle}

\usepackage{lineno,hyperref}
\modulolinenumbers[5]

\usepackage{amssymb}
\usepackage{enumerate}
\usepackage{multirow}
\usepackage{amsmath}

\usepackage{array}
\usepackage{algorithmic}
\usepackage{booktabs}

\usepackage{threeparttable}

\newcommand{\tabincell}[2]{\begin{tabular}{@{}#1@{}}#2\end{tabular}}

\usepackage{url}
\usepackage{xcolor}
\definecolor{newcolor}{rgb}{.8,.349,.1}

\begin{document}

\begin{frontmatter}

\title{Heterogeneous Separation Consistency Training for Adaptation of Unsupervised Speech Separation}

\author{Jiangyu Han}
\author{Yanhua  Long \corref{cor1}} 
\cortext[cor1]{Corresponding: yanhua@shnu.edu.cn}

\address{Shanghai Normal University, Shanghai, China}

\begin{abstract}
Recently, supervised speech separation has made great progress.
However, limited by the nature of supervised training, most existing separation
methods require ground-truth sources and are trained on synthetic datasets.
This ground-truth reliance is problematic, because the ground-truth signals
are usually unavailable in real conditions.
Moreover, in many industry scenarios, the real acoustic characteristics
deviate far from the ones in simulated datasets. Therefore,
the performance usually degrades significantly when applying
the supervised speech separation models to real applications.
To address these problems, in this study, we propose a novel separation
consistency training, termed SCT, to exploit the real-world unlabeled mixtures
for improving cross-domain unsupervised speech separation in an iterative manner, by
leveraging upon the complementary information obtained from heterogeneous (structurally distinct but behaviorally complementary) models.
SCT follows a framework using two heterogeneous neural networks (HNNs) to produce
high confidence pseudo labels of unlabeled real speech mixtures. These
labels are then updated, and used to refine the HNNs to produce more reliable consistent separation results for real mixture pseudo-labeling.
To maximally utilize the large complementary information between different
separation networks, a cross-knowledge adaptation is further proposed. Together with simulated dataset, those
real mixtures with high confidence pseudo labels are then
used to update the HNN separation models iteratively. In addition,
we find that combing the heterogeneous separation outputs by a simple
linear fusion can further slightly improve the final system performance.
The proposed SCT is evaluated on both public reverberant English and
anechoic Mandarin cross-domain separation tasks.
Results show that, without any available ground-truth of target
domain mixtures, the SCT can still significantly outperform our two
strong baselines with up to 1.61 dB and 3.44 dB scale-invariant signal-to-noise ratio (SI-SNR) improvements,
on the English and Mandarin cross-domain conditions respectively.  
\end{abstract}

\end{frontmatter}

% \linenumbers

%% main text
\section{Introduction}
\label{sec:intro}

Multi-speaker interaction scenarios are very common in real-world speech processing applications. Speech separation,
separating each source signal from mixed speech, is one of
the most important technology for these applications, including
speaker diarization, speaker verification, multi-talker speech
recognition, etc.

Because of the importance of speech separation, numerous studies have focused on this topic, including the traditional
time-frequency (T-F) domain separation methods \cite{dpcl, yu2017permutation, kolbaek2017multitalker, chen2017deep, luo2018speaker, wang2018alternative, wang2018supervised,williamson2015complex, tan2019complex, choi2018phase, dccrn, fu2021desnet}, and the recent popular time-domain approaches
\cite{tasnet, dprnn, zhang2020end, dptnet, sepformer, cd1, cd2, han2020attention}.
All these contributions have led to significant progress on the
single-channel speech separation. Most of them follow a mask learning
pattern that aims to learn a weighting matrix (mask) to capture
relationship between the isolated clean sources. The mask is then used to
separate each source signal with an element-by-element multiplication.
In addition, some researchers also concentrate on
learning clean sources directly from the mixed speech,
which is known as mapping-based separation \cite{chen2020synthesis, denseunet, luo2021rethinking}.

Reviewing recent speech separation techniques, most of them
are supervised ones with their own advantages. 
Such as, the T-F domain methods take spectrogram as input features 
that are good at capturing the phonetic structure
of speech \cite{yang2019improved};
the time-domain methods pay more attention to the fine
structure of speech but are vulnerable to environmental 
or background variations;
the masking-based methods are effective
in utilizing the clean speech of training corpus but 
sensitive to speech with signal-to-noise ratio (SNR)
variations; the mapping-based methods show
more robustness for tasks with a wide range of SNR \cite{zhang2016deep}, etc.
To fully exploit advantages over different approaches,
some studies focus on integrating different approaches into an ensemble training framework.
For example, authors in \cite{yang2019improved} constructed a time-and-frequency
feature map by concatenating both time and time-frequency domain acoustic features
to improve separation performance. 
For improving the singing voice extraction, in \cite{shi2020singing}, several attention-based fusion strategies
were proposed to utilize the complementarities between masking and mapping
spectrograms using a minimum difference masks (MDMs) \cite{shi2020spectrograms}
criterion.

Although the supervised speech separation methods or their
combinations have performed well on data with the same or similar acoustic properties
as the simulation training sets, the performance on cross-domain
real-world mixtures is still quite poor. The main problem of supervised training
is the strong reliance on individual ground-truth source signals.
It heavily precludes technique scaling to widely available real-world mixtures,
and limits progress on wide-domain coverage speech separation tasks.
In real scenarios, the isolated sources are difficult to collect,
because they are usually contaminated by cross-talk and
unknown acoustic channel impulse responses.
Therefore, it's very difficult to provide golden-standard handcrafted labels for a large number
of real-world mixtures to train a supervised separation system from scratch.
Moreover, adapting a well pre-trained supervised system to target
real acoustics is also challenging, because the distribution of
sound types and reverberation may be unknown and hard to estimate.

One approach to improve real-world unsupervised speech separation is to
directly use the real acoustics in system training. To this end,
some latest works start to separate speech from unsupervised or semi-supervised perspectives.
In \cite{mixit,waspaa21, sivaraman2021adapting}, a mixture invariant training (MixIT) that
requires only single-channel real acoustic mixtures was proposed.
MixIT uses mixtures of mixtures (MoMs) as input, and sums over
estimated sources to match the target mixtures instead of the
single-source references. As the model is trained to separate the MOMs
into a variable number of latent sources, the separated sources can be
remixed to approximate the original mixtures.
Motivated by MixIT, authors in \cite{zhang2021teacher} proposed a
teacher-student MixIT (TS-MixIT) to alleviate the over-separation problem
in the original MixIT. It takes the unsupervised model trained by MixIT
as a teacher model, then the estimated sources are filtered and selected
as pseudo-targets to further train a student model using standard
permutation invariant training (PIT) \cite{kolbaek2017multitalker}.
Besides, there are other unsupervised separation attempts as well,
such as the co-separation \cite{gao2019co}, adversarial unmix-and-remix \cite{hoshen2019towards},
and Mixup-Breakdown \cite{lam2020mixup}. All these recent efforts 
indicate how to well exploit the real-world unlabeled mixtures to
boost the current separation systems becomes very fundamental,
important, and challenging.

In this study, we also focus on improving the unsupervised speech separation,
a novel speech separation adaptation framework, termed separation
consistency training (SCT),
is proposed. Different from previous works,  SCT aims to
leverage the separation consistency between heterogeneous separation
networks to produce high confidence pseudo-labels of unlabeled
acoustic mixtures. These labels and networks are updated iteratively
using a cross-knowledge adaptation approach to achieve more accurate
pseudo-labels and better target speech separation models.
In SCT, two separation networks with a heterogeneous structure are
used, one is the current popular masking-based time-domain speech
separation model, Conv-TasNet\cite{tasnet},
and the other is our recent
proposed mapping-based time-frequency domain separation model, DPCCN \cite{dpccn}.
These two networks are then used to generate consistent separation
results for target domain unlabeled mixture labeling.
The advantages behind using heterogeneous networks instead of homogeneous
ones are that, besides the mixture labeling, the complementary information
between these heterogeneous models
is expected to attain large diversity for label creation.
By doing so,
it provides an increased chance to produce and select more
informative target mixtures as iterative training samples that a single source separation model could not produce by itself.
In addition, a simple linear fusion strategy is proposed to combine the heterogeneous separation outputs to further improve the final separation performance.

Our experiments are performed on three open-source datasets,
the anechoic English Libri2Mix \cite{libri2mix} is taken as the source domain data,
the reverberant English WHAMR! \cite{whamr}
and anechoic Mandarin Aishell2Mix \cite{dpccn} are our two target domain datasets.
Extensive results show that, the proposed SCT is very effective
to improve the unsupervised cross-domain speech separation performance.
It can significantly outperform two strong baselines with up to
1.61 dB and 3.44 dB
scale-invariant signal-to-noise ratio (SI-SNR)
\cite{sisnr} improvements on the English and Mandarin cross-domain
tasks, respectively. Besides, we find that, our separation
consistency selection can achieve competitive performance
with the data selection using ground-truth sources as references during
the target heterogeneous model adaptation.

\section{Previous work}
\label{sec:2}

\subsection{Conv-TasNet}
\label{subsec:convtas}

Conv-TasNet is a time-domain, masking-based speech separation technique that
proposed in \cite{tasnet}. Compared with most time-frequency domain algorithms, 
Conv-TasNet shows superior separation performance
on the standard public WSJ0-2mix \cite{dpcl} dataset, and has become the 
mainstream speech separation approach. This network has attracted
widespread attention and been further improved in many recent works
\cite{demystifying, sonning2020performance, ochiai2020beam, von2020end}.

Conv-TasNet consists of three parts: an encoder (1d convolution layer),
a mask estimator (several convolution blocks), and a decoder (1d deconvolution layer).
The waveform mixture is first encoded by the encoder and then is
fed into the temporal convolutional network (TCN)
\cite{lea2016temporal, lea2017temporal, bai2018empirical} based mask
estimator to estimate a multiplicative masking function
for each source. Finally, the source waveforms are reconstructed by
transforming the masked encoder representations using the decoder.
More details can be found in \cite{tasnet}.

\subsection{DPCCN}
\label{subsec:dpccn}

DPCCN is our recent work in \cite{dpccn}, it is a time-frequency domain,
mapping-based speech separation technique. Results in \cite{dpccn} show that
DPCCN can achieve much better performance and
robustness over other state-of-the-art separation methods in 
environmental complicated conditions.

DPCCN follows a U-Net \cite{unet} style to  encode the mixture spectrum into a
high-level representation, then decodes it into the clean speech.
In DPCCN, DenseNet \cite{densenet} is used to alleviate the vanishing-gradient
problem and encourage the feature reuse; TCN is clamped between the
codec to leverage long-range time information; A pyramid pooling layer
\cite{pyramid} is introduced after decoder to improve its global modeling
ability. The detailed information can be found in \cite{dpccn}.

%\begin{figure}[ht]
%  \centering
%  %\setlength{\abovecaptionskip}{0.cm}
%  \includegraphics[width=10.5cm]{dpccn.png}
%  \caption{The structure of DPCCN network. The skip connection means concatenation operation. Each 2-D convolutional block includes a 2-D convolution, exponential linear units (ELU), and instance normalization (IN). Each 2-D deconvolutional block includes a 2-D deconvolution, ELU and IN. The TCN blocks contain 2 layers TCN, and each includes 10 TCN blocks composed by IN, ELU and 1-D convolution with dilation factors $1, 2, 4, ..., 512$.}
%  \label{fig:dpccn}
%  %\vspace{-0.5cm}
%\end{figure}

\section{Heterogeneous Separation Consistency Training}
\label{sec:sct}

The proposed separation consistency training is performed
on two different separation networks with heterogeneous structure.
In this section, we first present the principle of our SCT, then
introduce three SCT variants and their differences, including
basic SCT and the cross-knowledge adaptation. Next, two main
algorithms, consistent pseudo-labeling and selection (CPS), and
heterogeneous knowledge fusion (HKF) in the proposed SCT are 
described in detail. For simplicity, here we only consider
the speech separation scenario with two-speaker overlapped speech.

\subsection{Separation Consistency Training}
\label{seubsec:sctt}

Our separation consistency training is specially designed
to improve the unsupervised speech separation where the target mixtures
deviate far from the training simulation dataset. It follows a
heterogeneous separation framework, to create and select informative data 
pairs with high confidence pseudo ground-truth,
for iteratively improving cross-domain speech separation by
adapting the source separation models to the target acoustic
environments. Because the whole framework heavily
relies on the consistent separation results of the unlabeled mixtures
and a separation consistency measure for pseudo-labeling, we call
the whole training process as separation consistency training (SCT).

\begin{figure}[t]
  \centering
  \includegraphics[width=8cm]{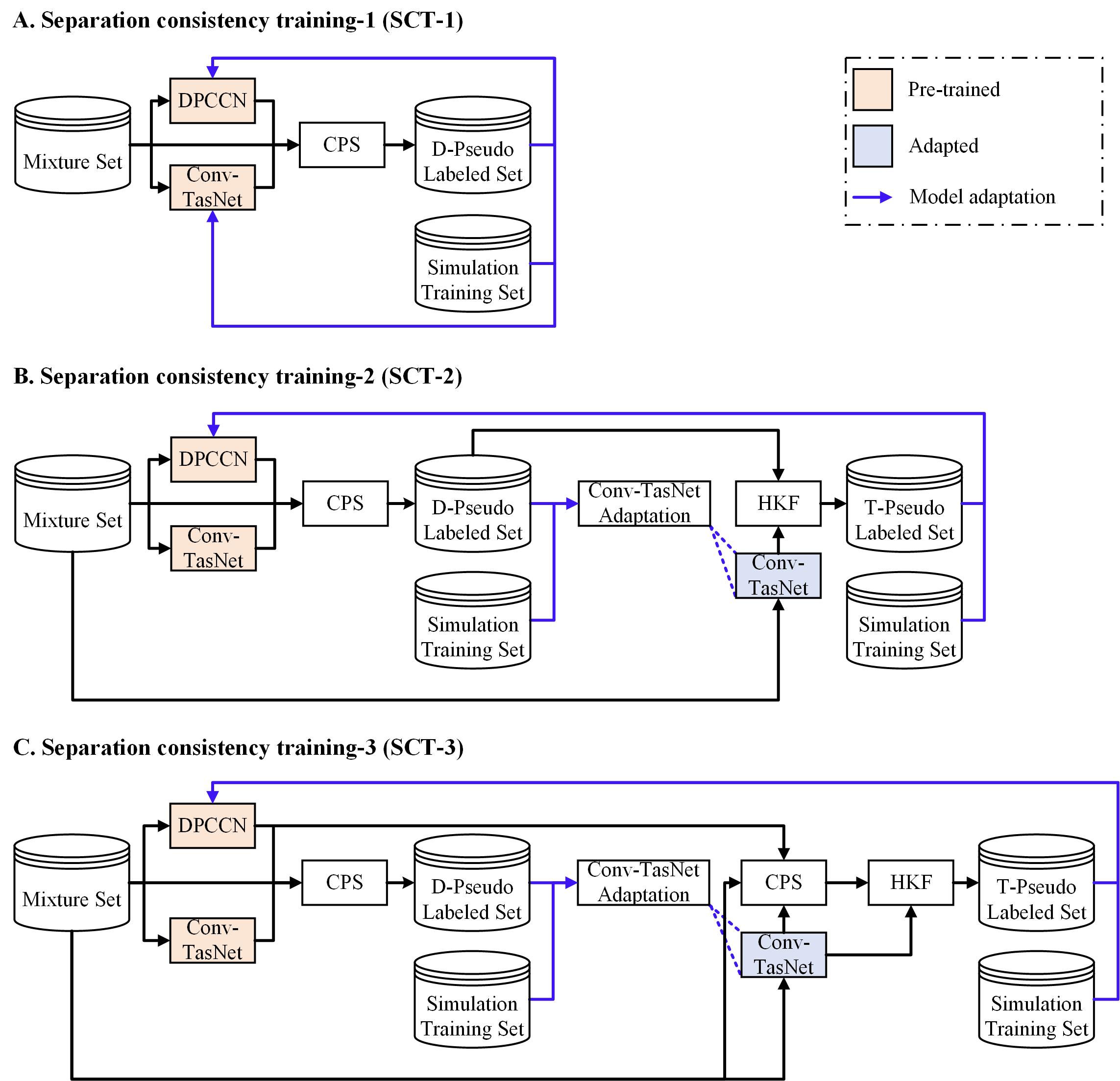}
    \caption{The flowcharts of three SCT variants for single iteration.
    (A) Framework of the 1st SCT variant (SCT-1). The selected DPCCN outputs with
    pseudo labels are used to update both Conv-TasNet and DPCCN. (B) Framework of the 2nd SCT variant (SCT-2) with
    the cross-knowledge adaptation. (C) Framework of the 3rd SCT variant (SCT-3).
    Two-stage CPS are used to refine the pseudo-labeling.
  }
  \label{fig:sct}
  %\vspace{-0.5cm}
\end{figure}

\textbf{Basic SCT.} Given a large amount of or even limited unlabeled target mixtures,
the basic SCT procedure can be divided into three main steps:
\begin{enumerate}
  \item[(a)] \textbf{Mixture separation}. Separate each unlabeled mixture using
        two heterogeneous separation models that have been
        well-trained on the source simulated training set;

  \item[(b)] \textbf{Consistent pseudo-labeling and selection (CPS).} Based on separation results in step (a),
  calculate a separation consistency measure (\verb"SCM", Eq.(\ref{eq:scm})) and a mixture
  separation consistency measure (\verb"mSCM", Eq.(\ref{eq:mscm})) to evaluate the confidence of separation outputs. Then,
  select those unlabeled mixtures with high consistent confidence and
  their corresponding separation results as pseudo ground-truth to form a ``Pseudo Labeled Set";

  \item[(c)] \textbf{Iterative model adaptation}. Combine the ``Pseudo Labeled Set" and the original
        source domain ``Simulation Training Set" together to refine
        the source models to learn the target domain acoustics. And then, repeat the above process in an iterative manner.

\end{enumerate}

The two separation models in step (a) usually have
comparable performance but with heterogeneous neural network
structures. The bigger difference between the models, the better complementary
information will be achieved. In this study, we choose  DPCCN and
Conv-TasNet that presented in Section \ref{sec:2} as the heterogeneous candidates.
The former is taken as the primary model, while the latter is regarded as a reviewer model.
Conv-TasNet \cite{tasnet} is the current popular masking-based time-domain separation
model, while DPCCN \cite{dpccn} is our recent proposed mapping-based time-frequency domain
model with good robustness to complicate acoustic environments.
The huge difference between Conv-TasNet and DPCCN  guarantees
the large diversity of the separated results. This
diversity provides an increased chance to improve source models
iteratively, because it can produce more informative
target mixtures as new iterative training samples that the primary source model
could not produce itself. Actually,
during CPS in step (b), each model in this SCT heterogeneous framework is 
a reviewer for the other, any input mixtures will be double inspected by
the two reviewers from different perspectives,
only those mixtures with consistent separation performance of both will be
selected. In this way, the double inspecting mechanism under a heterogeneous
framework ensures the high confidence of pseudo ground-truth for
each selected mixture in
the target domain.

The whole framework of above basic SCT is demonstrated in
the first variant of our proposed SCT, subfigure (A) (SCT-1) of
Fig.\ref{fig:sct}. In SCT-1, the detail of consistent pseudo-labeling and selection (CPS)
is presented in the next section, Section \ref{ssec:scs}, and illustrated in
Fig.\ref{fig:scs}(A).
``D-Pseudo Labeled Set" (``D-'' means DPCCN's outputs) contains the data pairs of selected unlabeled mixtures and
their pseudo ground-truth that derive from
the individual separation outputs of the primary model DPCCN.
Together with the original source domain ``Simulation Training Set'', 
both the primary and reviewer models are refined and adapted
to the target domain in each
single iteration.
It is worth noting that the model adaptation with the combined training set is necessary for SCT algorithm. As our source models have been trained well on the simulation set, and the pseudo ground-truth of ``D-Pseudo Labeled Set" is actually generated by DPCCN, which means if we only use the simulation set or ``D-Pseudo Labeled Set" to adapt the primary source model, DPCCN, the training gradient will be very small or even 0. In this case, the error between model outputs and labels is difficult to back propagate and the adaptation process will fail. However, if we adapt model using both ``Simulation Training Set'' and  ``D-Pseudo Labeled Set'',
although the error between model outputs and ground-truth is small, the model is still can be adapted to the target domain. For example, a simple neural network can be depicted as $\bf{y}=w*x + b$ , where $\bf{y}, w, x, b$ are model output, parameter weight, model input, and parameter bias, respectively. The partial differential to the weight $\bf{w}$ is model input $\bf{x}$. Back to our scenario, by combining ``Simulation Training Set'' and  ``D-Pseudo Labeled Set'',
the target domain data can engage in the model adaptation with the loss of the source domain simulation set.

\textbf{SCT with cross-knowledge adaptation.} To fully exploit the
complementary information between heterogeneous networks,
a cross-knowledge adaptation is proposed to improve the basic SCT.
The framework is illustrated in the 2nd variant of SCT (SCT-2) in
Fig.\ref{fig:sct}(B). Different from  basic SCT,
in SCT-2, the reviewer Conv-TasNet is first updated using the
combined ``D-Pseudo Labeled Set" and 
``Simulation Training Set", i.e., the pseudo ground-truth
of the primary model is used to guide the reviewer model's
adaptation. Next, we re-separate all the unlabeled mixtures
using the updated reviewer to achieve more accurate separation
outputs. Then, all the pseudo ground-truth in ``D-Pseudo Labeled Set"
are replaced by the corresponding new individual outputs that produced by the
updated reviewer Conv-TasNet to construct a new pseudo  labeled set ``T-Pseudo Labeled Set" (``T-'' means Conv-TasNet's outputs).
Finally, the ``T-Pseudo Labeled Set" and ``Simulation Training Set"
are combined together to adapt the primary model DPCCN as in SCT-1.
In this model adaptation, the pseudo ground-truth
of the reviewer model is used to supervise the primary model training.
Just like the teacher-student learning, in the whole SCT-2,
the primary and reviewer model can benefit each other,
the learned knowledge of them is cross-used
as a guide to improve the target model adaptation. Therefore,
we call this adaptation procedure as ``cross-knowledge adaptation"
for simplicity. In addition, as the
``T-Pseudo Labeled Set" is actually a combination of prior selected
separation consistency statistics in ``D-Pseudo Labeled Set"  and the
new pseudo ground-truth from updated Conv-TasNet, thus, in Fig.\ref{fig:sct}, we use
the heterogeneous knowledge fusion (HKF) block to represent
this knowledge combination.
Details of HKF are demonstrated in subfigure (D) of Fig.\ref{fig:scs} and
Section \ref{ssec:hkf}.

Subfigure (C) (SCT-3) of Fig.\ref{fig:sct} is a variant of SCT-2 with
minor modification before HKF block. In SCT-3, the
CPS is performed twice.
The first CPS is performed as the same in SCT-1 and SCT-2, while in the
second CPS, the separation consistency statistics, \verb"SCM"
and \verb"mSCM" are re-computed and updated using
both mixture separation outputs of DPCCN and the updated Conv-TasNet.
Other operations are all the same as in SCT-2.
The motivation behind this two-stage CPS is that, the adapted Conv-TasNet
can provide more accurate separation results of target domain mixtures,
which makes the second stage CPS produce
more reliable consistent separation results for unlabeled
mixture pseudo-labeling in each SCT iteration.

In summary, in this section, we present three variants of the proposed
SCT, one is the basic structure, and the others are two enhanced
SCT variants with cross-knowledge adaptation. Details of the CPS and HKF
blocks used in SCT are described in the following sections.

\subsection{Consistent Pseudo-labeling and Selection}
\label{ssec:scs}

\begin{figure*}[t]
  \centering
  \includegraphics[width=13cm]{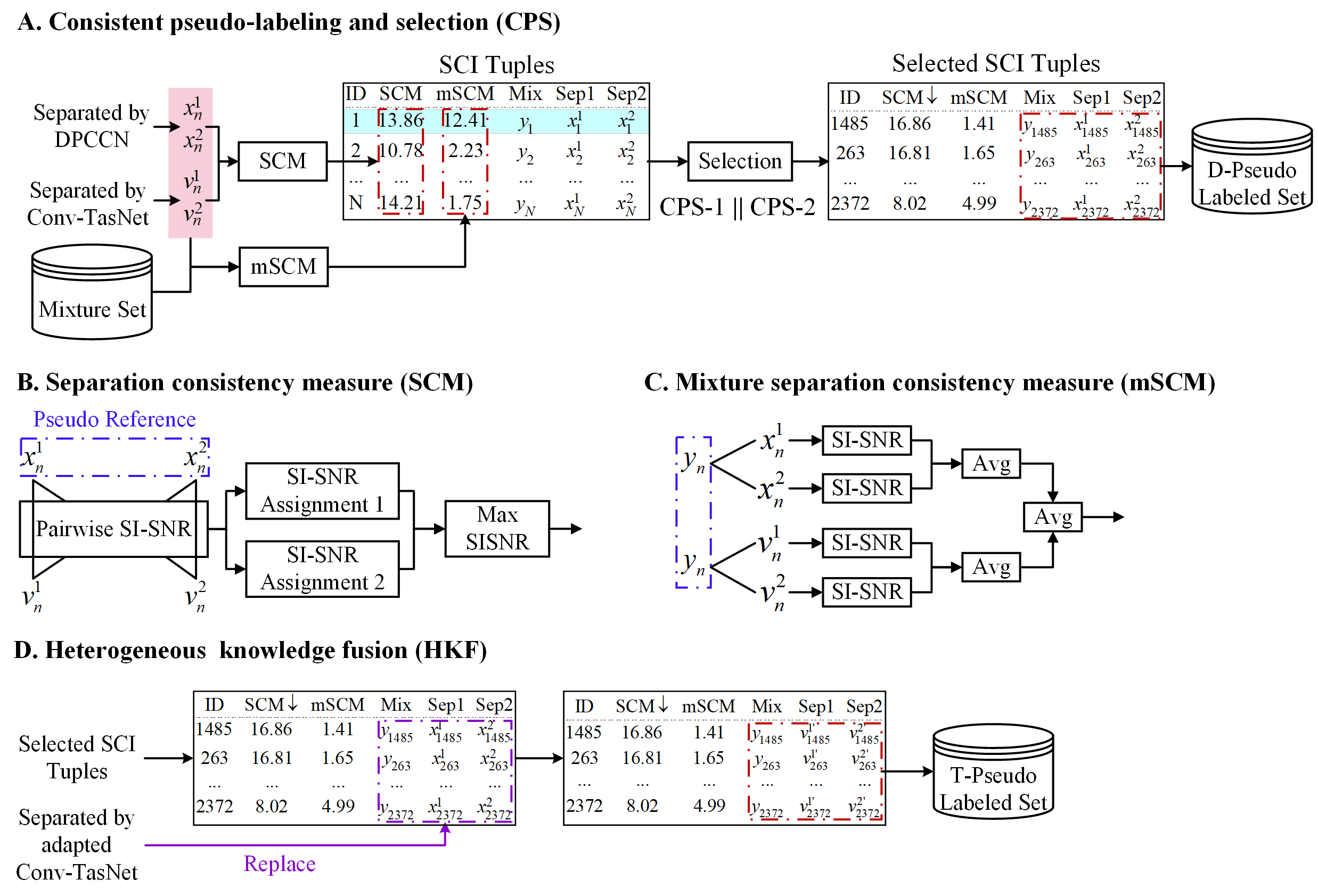}
  \caption{(A) The whole framework of consistent pseudo-labeling and selection (CPS).
  (B) The flowchart of separation consistency measure. (C) The flowchart of the mixture separation consistency measure. (D) The heterogeneous knowledge fusion.}
  \label{fig:scs}
  %\vspace{-0.5cm}
\end{figure*}

The consistent pseudo-labeling and selection (CPS) block in the proposed
SCT aims to produce high quality individual pseudo ground-truth of each
unlabeled mixture based on the outputs of two heterogeneous networks
and the original mixture speech. The whole CPS procedure is illustrated
Fig.\ref{fig:scs}(A). It contains two main stages, the first one is the confidence measure calculation, follows by the pseudo ground-truth
selection.

\textbf{Confidence measure calculation.} Two measures are calculated
in this stage, one is separation consistency measure (\verb"SCM", Eq.(\ref{eq:scm})), and
the other is mixture separation consistency measure (\verb"mSCM", Eq.(\ref{eq:mscm})). Both
of them are used to evaluate the confidence between heterogeneous separation outputs that produced by DPCCN and Conv-TasNet.

As shown in the left part of Fig.\ref{fig:scs}(A), given
$N$ unsupervised mixed speech with each contains $M$ single
sources, here we assume $M=2$. For the $n$-th mixture, the \verb"SCM" is
calculated by taking the individual separation output $\mathbf{x}_n$ of the primary
model DPCCN as pseudo reference as follows:
\begin{equation}
\label{eq:scm}
{\verb"SCM"}\left(\mathbf{x}_n, \mathbf{v}_n\right) =
\max_{\mathbf{P}} \frac{1}{M} \sum_{i=1}^M{\mathrm{\operatorname{SI-SNR}}}\left(x_n^i, [\mathbf{Pv}_n]_i\right)
\end{equation}
where $\mathbf{x}_n = [x_n^1, x_n^2, ..., x_n^M]^{\rm{T}}$,
$\mathbf{v}_n = [v_n^1, v_n^2, ..., v_n^M]^{\rm{T}}$ are the
$M$ individual separation speech signals that separated by DPCCN and
Conv-TasNet for the $n$-th input mixture, respectively.
$x_n^i$ and $v_n^i$ are the $i$-th individual signal. $\mathbf{P}$ is an
$M \times M$ permutation matrix, $[\cdot]_i$ denotes selecting $i$-th element
from the matrix, and $\rm{T}$ is the operation of transpose.
The SI-SNR in Eq.(\ref{eq:scm}) is the standard
scale-invariant signal-to-noise ratio (SI-SNR)\cite{sisnr} that
used to measure the performance of state-of-the-art speech separation
systems. It is defined as:
\begin{equation}
\label{eq:sisnr}
\mathrm{\operatorname{SI-SNR}}(s, \hat{s}) = 10\log_{10}\left(\frac{\lVert{\frac{\langle \hat{s}, s \rangle}{\langle s, s \rangle}s}\rVert^2}{\lVert{\frac{\langle \hat{s}, s \rangle}{\langle s, s \rangle}s} - \hat{s} \rVert^2}\right)
\end{equation}
where $s$ and $\hat{s}$ are the reference and estimated speech, respectively.
$\lVert \cdot \rVert^2$ denotes the signal power,
$\langle \cdot \rangle$ is the inner-product operation.

Fig.\ref{fig:scs}(B) shows a two-speaker \verb"SCM" process for the $n$-th mixture. The DPCCN outputs,
$x_n^1$, $x_n^2$ are taken as references to calculate the pairwise SI-SNR with the
Conv-TasNet outputs, $v_n^1$ and $v_n^2$. In this case,
there are two permutation combinations, namely $[x_n^1, v_n^1; x_n^2, v_n^2]$
and $[x_n^1, v_n^2; x_n^2, v_n^1]$, then \verb"SCM" compares the averaging pairwise SI-SNR for each assignment
and takes the highest value to represent the separation consistency between
two heterogeneous networks outputs. The higher \verb"SCM", the better consistency of unlabeled separation outputs we can trust. However,
when the input mixtures are hard to separate for both heterogeneous
networks, $\mathbf{x}_n$ and $\mathbf{v}_n$ can be very close to
the original mixture speech, and they could also result in a very
high \verb"SCM". 
In this case, the pseudo reference $\mathbf{x}_n$ may be far
from the ground-truth and may not be qualified for the source model adaptation.
To alleviate this situation, the following \verb"mSCM" is introduced from another perspective
to evaluate the quality of target domain mixture separation results and enhance the confidence of selected results.

The mixture separation consistency measure (\verb"mSCM"), aims to
measure the consistency between the outputs of heterogeneous networks
and the original input mixture $y_n$. It is defined as:
\begin{equation}
\label{eq:mscm}
{\verb"mSCM"}(y_n, \mathbf{x}_n, \mathbf{v}_n) = \frac{1}{2M} \sum_{i=1}^M \sum_{\phi} {\mathrm{\operatorname{SI-SNR}}}(y_n, \phi_n^i)
\end{equation}
where $\phi _n^i \in \{x_n^i, v_n^i\}$ is the $i$-th individual output of
DPCCN or Conv-TasNet of $n$-th input mixture as shown in Eq.(\ref{eq:scm}).
Fig.\ref{fig:scs}(C) gives a detailed operation of \verb"mSCM" under a two-speaker
case, and as shown in Eq.(\ref{eq:mscm}), we see the
average SI-SNR between the input mixture and
all separated outputs are calculated. Different from \verb"SCM", the \verb"mSCM"
evaluate the confidence of separation results in an opposite way and the lower is desired.
We believe that, in most conditions, the waveform of  well-separated results should be very different from the original mixture. Therefore, the corresponding \verb"mSCM" will be in a low position. 
It is noted that when the input mixture has a high input SNR, the lower \verb"mSCM" constraint will filter out its separated results. 
Even though, the lower \verb"mSCM" hypothesis still makes sense, because the filtered speech with high input SNR is somehow homogeneous and has limited benefits to model adaptation. In addition, the high input SNR cases are rare for cross-domain task. Therefore, the lower \verb"mSCM" constraint is safe and effective in most conditions. 

\textbf{Pseudo ground-truth selection. } After computing both \verb"SCM" and \verb"mSCM"
statistics of input mixtures, we re-organize all the statistics and speech
signals that related to each unlabeled input mixture in a new
data tuple format to facilitate the pseudo ground-truth selection.
As shown in Fig.\ref{fig:scs}(A), we call each data tuple as a
``separation consistency information (\verb"SCI")" tuple, and it is organized
as:
\begin{equation}
 {\verb"SCI"}= \{  \verb"ID, SCM, mSCM, Mix, Sep1, Sep2"\}
\end{equation}
where \verb"ID" is the  mixture ID, \verb"Mix" is the input mixture speech signal,
\verb"Sep1" and \verb"Sep2" are the two individual speech signals that separated by DPCCN.
With these \verb"SCI" tuples, we then perform the pseudo ground-truth selection in two ways:
\begin{itemize}
  \item CPS-1: Select \verb"SCI" pairs with \verb"SCM" value lies in the top $p\%$ \verb"SCM" range, $p\in [0,100]$.
  \item CPS-2: Select \verb"SCI" tuples with the following constraint:
  \begin{equation}
   \verb"SCI"_s = \left\{ \verb"SCI"_k \mid (\verb"SCM"_k > \alpha) \cap (\verb"mSCM"_k < \beta ) \right\}
  \end{equation}
  where $k = 1,2,...,N$, $\verb"SCI"_s$ and $\verb"SCI"_k$ are the selected \verb"SCI" tuples and
  $k$-th \verb"SCI" tuple, respectively. $\alpha$, $\beta$ are thresholds for \verb"SCM" and \verb"mSCM", respectively.
\end{itemize}

No matter for CPS-1 or CPS-2, the separated signals, \verb"Sep1" and \verb"Sep2", in all the selected
\verb"SCI" tuples will be taken as the high confidence
pseudo ground-truth for their corresponding mixture \verb"Mix". Then
the selected mixtures with pseudo ground-truths are taken to form the
``D-Pseudo Labeled Set" (pseudo ground-truth that produced by DPCCN)
for further separation model adaptation.  As discussed in
the definition of \verb"mSCM", compared with CPS-1, perhaps CPS-2 is better at dealing with
the difficult separation cases to some extent.

\subsection{Heterogeneous Knowledge Fusion}
\label{ssec:hkf}

The heterogeneous knowledge fusion (HKF), illustrated in Fig.\ref{fig:scs}(D)
is used during the cross-knowledge adaptation in SCT-2 and SCT-3.
HKF is a very simple operation just by replacing \verb"Sep1" and
\verb"Sep2" in the selected \verb"SCI" tuples of Fig.\ref{fig:scs}(A) with
the outputs of the adapted Conv-TasNet as in SCT-2 and SCT-3.
We use $v_n^{i'}$ to represent the $i$-th individual signal of $n$-th mixture separated by the adapted Conv-TasNet.
The updated new data tuples $\{\verb"Mix, Sep1, Sep2"\}$ are then picked to form the ``T-Pseudo Labeled Set" (pseudo ground-truths that produced by Conv-TasNet).
By doing so, the complementary information between
the prior knowledge of separation consistency information
that captured in the CPS block and the adapted Conv-TasNet are
subtly integrated to further refine the primary DPCCN.

\section{Experimental Setups}
\label{sec:exp}

\subsection{Dataset}
\label{subsec:dataset}

The publicly available English Libri2Mix \cite{libri2mix} is used as our
source domain dataset. Libri2Mix is a recent released anechoic separation corpus that contains artificial mixed speech from Librispeech \cite{librispeech}.
We use the Libri2Mix generated from ``train-100" subset to train our models.
Two target domain datasets are used to validate our proposed methods, one is
the English WHAMR! \cite{whamr}, the other is the Mandarin Aishell2Mix \cite{dpccn}.
WHAMR! is a noisy and reverberant version of the WSJ0-2mix dataset \cite{dpcl} with
four conditions (clean and anechoic, noisy and anechoic, clean and reverberant,
noisy and reverberant). We take the clean and reverberant condition
to evaluate the cross-domain speech separation performance.
Note that the evaluation references of WHAMR! are also
reverberant rather than anechoic. Aishell2Mix is created by ourselves \cite{dpccn},
it is anechoic and released in \cite{git_data}. Each mixture in Aishell2Mix is
generated by mixing two-speaker utterances from Aishell-1 \cite{aishell}.
These utterances are randomly clamped to 4 seconds and rescaled to a random
relative SNR between 0 and 5 dB. All datasets used in this study are resampled to 8kHz.
The mixtures in both target domain datasets, WHAMR! and Aishell2Mix,
are taken as the real-world unlabeled speech.
Only the ground-truth of test sets in WHAMR! and Aishell2Mix are available
for evaluating the speech separation performance, the training and development
sets are all unlabeled. More details can be found in Table \ref{tab:data}.
It is worth noting that,
the target domain development sets used to
supervise the model adaptation are also with pseudo ground-truth that
produced by our proposed SCT.

\begin{table}[!htbp]
\footnotesize
 \caption{Cross-domain dataset information.}
  \label{tab:data}
 \setlength{\tabcolsep}{1.15mm}
  \centering
  \begin{threeparttable}
  \begin{tabular}{c|c|c|c|c|c|c}
    \hline
	Dataset & Acoustics & Type & \#Spks & \#Utts & Hours & Oracle  \\
    \hline
    \multirow{3}{*}{\tabincell{c}{Libri2Mix\\(Source)}} &
    \multirow{3}{*}{English/A} & train & 251 & 13900 & 58 & \checkmark  \\
         & & dev & 40 & 3000 & 11 & \checkmark \\
         & & test & 40 & 3000 & 11 & \checkmark \\
    \hline
    \multirow{3}{*}{\tabincell{c}{WHAMR!\\(Target)}} &
    \multirow{3}{*}{English/R} & train & $101^*$  & 20000 & 30 & -  \\
         & & dev & $101^*$ & 5000 & 10 & - \\
         & & test & 18 & 3000 & 5 & \checkmark  \\
    \hline
    \multirow{3}{*}{\tabincell{c}{Aishell2Mix\\(Target)}}  & \multirow{3}{*}{Mandarin/A} & train & 340 & 10000 & 11 & - \\
     &   & dev & 40 & 3000 & 3 & - \\
     &   & test & 20 & 3000 & 3 & \checkmark \\
  	\hline
  \end{tabular}
  \begin{tablenotes}
    \footnotesize
    \item  ``A'' and ``R'' refer to anechoic and reverberant, respectively.
  ``Oracle" indicates whether the oracle (ground-truth)
  data is available. $``*"$ means the speakers of different sets are same.
  \end{tablenotes}
  \end{threeparttable}
\end{table}

\subsection{Configurations}
\label{subsec:config}

We keep the same network configurations of Conv-TasNet and DPCCN as in
\cite{tasnet} and \cite{dpccn}, respectively.
The model parameters of Conv-TasNet and DPCCN are 8.8M\footnote{ https://arxiv.org/pdf/1809.07454v1.pdf}  and 6.3M. When processing a 4-second speech, the number of multiply-accumulate (MAC) operations \cite{whitehead2011precision} of Conv-TasNet and DPCCN are 28.2G and 33.1G, which are evaluated using open-source toolbox\cite{opcounter}.
All models are trained with 100 epochs on 4-second speech segments.
The initial learning rate is set to 0.001 and halved if the accuracy of
development set is not improved in 3 consecutive epochs.
Adam \cite{adam} is used as the optimizer and the early stopping is applied
for 6 consecutive epochs. We use the standard negative SI-SNR \cite{sisnr}
as loss function to train all separation systems.
Utterance-level permutation invariant training (uPIT) \cite{kolbaek2017multitalker}
is used to address the source permutation problem.
All source model adaptation related experiments are finished within 20 epochs.
A Pytorch implementation of our DPCCN system can be found in \cite{git_dpccn}.

\subsection{Evaluation Metrics}
\label{subsec:eva}

As our task is to improve cross-domain unsupervised speech separation,
the performance improvement over the original mixture is more meaningful.
Therefore, we report the  well-known signal-to-distortion ratio improvement (SDRi)
\cite{vincent2006performance} and scale-invariant signal-to-noise
ratio improvement (SI-SNRi) \cite{sisnr} to evaluate our proposed method.

\section{Results and analysis}
\label{sec:results}

\subsection{Cross-domain Baselines}
\label{subsec:cdb}

\textbf{Baselines}. Both Conv-TasNet and DPCCN are taken as our cross-domain
baseline systems. Performance is evaluated on all the in-domain
Libri2Mix, and cross-domain WHAMR! and Aishell2Mix test sets.
Results are shown in Table \ref{tab:cross_domain}, where all separation
systems are trained only on the Libri2Mix.

\begin{table}[!htbp]
 \renewcommand\arraystretch{1.25}
  \caption{SDRi/SI-SNRi (dB) performance of Conv-TasNet and DPCCN on Libri2Mix, WHAMR!, and Aishell2Mix test sets. Systems are all trained on Libri2Mix.}
  \label{tab:cross_domain}
  \centering
  \scalebox{0.96}{
  %\begin{threeparttable}
  \setlength{\tabcolsep}{0.5mm}
  \begin{tabular}{l|ccc}
    \hline
	\multirow{2}{*}{System} & \multicolumn{3}{c}{SDRi/SI-SNRi (dB)} \\
	& Libri2Mix & WHAMR! & Aishell2Mix \\ 
  	\hline
  	Conv-TasNet & 12.41 / 11.98 & 6.83 / 6.45 & 2.57 / 2.08 \\
  	DPCCN & 13.48 / 13.04 & 8.99 / 8.50 & 5.78 / 5.09 \\
  	\hline
  \end{tabular}}
\end{table}

From Table \ref{tab:cross_domain}, three findings are observed:

\begin{enumerate}
  \item [1)] Compared with the performance on the in-domain Libri2Mix
             test set, there are huge cross-domain performance gaps exist on both the English and Mandarin target domain datasets.
  \item [2)] Separation performance degradation caused by the language
             mismatch is much more severe than the acoustic reverberation.
  \item [3)] DPCCN always shows much better speech separation performance than Conv-TasNet
             under both in-domain and cross-domain conditions.
\end{enumerate}

The first two findings confirm that the current speech separation systems are very
sensitive to cross-domain conditions, either for the time-domain Conv-TasNet, or
the time-frequency domain DPCCN. The third observation shows 
the better system robustness of
DPCCN over Conv-TasNet. We believe that the robustness gain of DPCCN mainly comes from using spectrogram to represent speech.
For complicated tasks, 
such a handcrafted signal representation can provide more stable speech features than network learning.
That's why we take the DPCCN individual outputs
as references to calculate \verb"SCM" for pseudo ground-truth selection
as described in Section \ref{ssec:scs}. We believe  more reliable separation
hypotheses can result in better pseudo ground-truth.

\begin{table}[!htbp]
 \renewcommand\arraystretch{1.25}
  \caption{Performance of Conv-TasNet and DPCCN trained with ground-truth labels on WHAMR!
            and Aishell2Mix test sets.}
  \label{tab:supervised}
  \centering
  \setlength{\tabcolsep}{0.9mm}
  \begin{tabular}{l|c|c|c}
    \hline
	System & Dataset &SDRi (dB) & SI-SNRi (dB) \\
  	\hline
    \multirow{2}{*}{Conv-TasNet} & WHAMR! & 11.03 & 10.59 \\
    & Aishell2Mix & 9.00 & 8.32 \\
    \hline
    \multirow{2}{*}{DPCCN} & WHAMR! & 11.01 & 10.56 \\
    & Aishell2Mix & 8.86 & 8.14 \\
  	\hline
  \end{tabular}
\end{table}

\textbf{Training with ground-truth labels.} For results comparison and analysis,
we also report the supervised separation performance  of Conv-TasNet and DPCCN that trained with ground-truth labels
in Table \ref{tab:supervised}, where all separation systems
are trained with in-domain ground-truth sources of WHAMR! and Aishell2Mix. Interestingly, on the reverberant WHAMR! dataset, DPCCN and
Conv-TasNet achieve almost the same results, while on the
Aishell2Mix, DPCCN performs slightly worse than the Conv-TasNet.
Coupled with the better cross-domain
separation behaviors in Table \ref{tab:cross_domain}, we
take the DPCCN as our primary system, and the Conv-TasNet as the
reviewer in all our following experiments.

\subsection{Performance Evaluation of SCT on Aishell2Mix}
\label{subsec:sgp}

From Table \ref{tab:cross_domain} baseline results, we see
the domain mismatch between English and Mandarin datasets
is much larger than the two different English datasets.
Therefore, in this section, we choose to first
examine the proposed SCT
on the Libri2Mix-Aishell2Mix (source-target) unsupervised
cross-domain task, including evaluating the consistent pseudo-labeling
and selection methods, CPS-1 and CPS-2, and different SCT variants
for unsupervised model adaptation. Then, the optimized SCT
is generalized to the WHAMR! dataset in Section \ref{subsec:esw}.

\begin{table}[!htbp]
  \caption{SDRi/SI-SNRi performance of DPCCN with CPS-1 and SCT-1 on Aishell2Mix test set.}
  \label{tab:scs-1}
  \renewcommand\arraystretch{1.25}
  \centering
 \begin{threeparttable}
  \setlength{\tabcolsep}{0.9mm}
  \begin{tabular}{l|c|c|c}
    \hline
    top $p\%$ \verb"SCM" & Adaptation & SDRi (dB) & SI-SNRi (dB) \\
    \hline
    $p=10\%$ & - & 5.24 & 4.49 \\
    $p=25\%$ & - & 5.65 & 5.05 \\
    $p=50\%$ & - & 5.66 & 5.11 \\
     & \checkmark & \bf{5.98} & \bf{5.32} \\
  	\hline
  \end{tabular}
  \begin{tablenotes}
  \footnotesize
  \item 
  ``-'' means training model from scratch with only pseudo labeled data. ``\checkmark'' means adapting model with the combined pseudo labeled data and the source domain Libri2Mix.
  \end{tablenotes}
\end{threeparttable}
\end{table}

\textbf{Initial examination of CPS-1.} The DPCCN performance of the first unlabeled mixture pseudo label
selection method, CPS-1, is first examined under SCT-1 framework in Table \ref{tab:scs-1}.
Results of line 1-3 are from DPCCN that trained from scratch using
CPS-1 outputs. These outputs are the ``D-Pseudo Labeled Set" in SCT-1
with top $p\%$ \verb"SCM" target domain Aishell2Mix data.
We find that the separation performance can be improved by increasing the pseudo labeled training mixtures. And when $p=50\%$, compared with the $p=25\%$ case,
the additional performance improvements are rather limited even  with an  additional $25\%$ data. Moreover,
results of the last line show that, instead of training
DPCCN from scratch, using the combined ``D-Pseudo Labeled Set" and ``Simulation Training Set" (Libri2Mix) to refine the source model (shown in Table \ref{tab:cross_domain}, SDRi/SI-SNRi are 5.78/5.09 respectively) can
further improve the target domain separation. 
In the following experiments,
we set $p=50\%$ for all the CPS-1 experiments, and use
Libri2Mix training set together with the ``Pseudo Labeled Set"
to fine-tune the source separation models for target model
adaptation.

\begin{table}[!htbp]
  \caption{SDRi/SI-SNRi (dB) performance of Conv-TasNet and DPCCN on Aishell2Mix test set under different SCT configurations.}
  \label{tab:sct_aishell}
  \begin{threeparttable}
  \centering
  \renewcommand\arraystretch{1.25}
  \setlength{\tabcolsep}{0.6mm}
  \begin{tabular}{c|c|c|c|c|c}
    \hline
	SCT  & System & \#Iter  & CPS-1 & CPS-2 & Oracle\\
	\hline
	\multirow{4}{*}{SCT-1} & \multirow{2}{*}{Conv-TasNet} & 1 &  5.14/4.63 & 5.47/4.90 & 5.98/5.39 \\
	& & 2 & 5.45/4.94 & 5.99/5.39 & 6.18/5.57 \\
	\cline{2-6}
	& \multirow{2}{*}{DPCCN} & 1 &  5.98/5.32 & 5.90/5.25 & 6.00/5.31 \\
	& & 2 &  6.17/5.50 & 6.03/5.39 & 6.10/5.44 \\
  	\hline
	\multirow{4}{*}{SCT-2} & \multirow{2}{*}{Conv-TasNet} & 1 &  5.14/4.63 & 5.47/4.90 & 5.98/5.39 \\
	& & 2 &  5.36/4.89 & \bf{6.15/5.52} & 6.21/5.65 \\
	\cline{2-6}
	& \multirow{2}{*}{DPCCN} & 1 &  6.05/5.52 & \bf{6.48/5.82} & 6.79/6.19 \\
	& & 2 &  5.49/5.05 & 6.43/5.81 & 6.45/5.91 \\
  	\hline
	\multirow{4}{*}{SCT-3} & \multirow{2}{*}{Conv-TasNet} & 1 &  5.14/4.63 & 5.47/4.90 & - \\
	& & 2 &  5.43/4.93 & 5.77/5.24 & - \\
	\cline{2-6}
	& \multirow{2}{*}{DPCCN} & 1 &  6.14/5.58 & 6.22/5.65 & - \\
	& & 2 &  6.02/5.52 & 6.10/5.56 & - \\
  	\hline  	
  \end{tabular}
  \begin{tablenotes}
  \footnotesize
  \item ``Oracle'' means using ground-truth as reference to calculate
  SI-SNR of separation outputs for selecting the pseudo ground-truth.
  All source models are well pre-trained on Libri2Mix. The best setup of $\{\alpha,\beta\}$
  in CPS-2 are $\{5,5\}$, $\{8,5\}$ in the 1st and 2nd
  iteration for all SCT variants, respectively. $\eta$ is set to 5 for
  ``Oracle selection''.
  \end{tablenotes}
\end{threeparttable}
\end{table}

\textbf{Evaluating SCT variants with both CPS-1 and CPS-2.}
Unlike only adapting DPCCN model as in the above CPS-1 initial
experiments, in Table \ref{tab:sct_aishell}, we present the performance
of both the updated target DPCCN and Conv-TasNet in each SCT iteration for
all the three types of SCT variants. Experiments
are still performed on the English-Mandarin cross-domain speech
separation task. All source models are pre-trained on the same supervised
Libri2Mix, then adapted to the Aishell2Mix condition using SCT-1 to SCT-3
frameworks separately. Besides the CPS-1 and CPS-2, in Table \ref{tab:sct_aishell},
we also report ``oracle selection" performance using ground-truth
as reference to calculate SI-SNR of separation outputs for selecting the pseudo ground-truth.
This ``oracle selection" performance can be taken as the upper bound of our pseudo-labeling with heterogenous neural network architecture.
Two oracle selection criterions are used in our experiments:
for SCT-1, we always calculate the best assignment SI-SNR between
DPCCN outputs and ground-truth, while for SCT-2 and SCT-3,
we use the SI-SNR scores between the ground-truth and DPCCN, Conv-TasNet
outputs separately to select their corresponding individual
separation signals as pseudo ground-truth, respectively.
The pseudo ground-truth selection threshold $\eta=5$ is unchanged
for each iteration in ``oracle selection". It is worth noting that,
the $\{\alpha,\beta,\eta\}$ are kept the same for both the
pseudo-labeling of unlabeled training and development datasets.

From the English-Mandarin cross-domain separation results in
Table \ref{tab:sct_aishell}, we can conclude the following observations:

\begin{enumerate}
    \item [1)] Overall performance: Compared with baselines in Table \ref{tab:cross_domain},
    the best SCT variant, SCT-2 with CPS-2, improves the unsupervised
    cross-domain separation performance significantly.
    Specifically, absolute 3.68/3.44 dB and 0.70/0.73 dB SDRi/SI-SNRi
    improvements are obtained for Conv-TasNet and DPCCN, respectively.
    Moreover, the best performance of SCT-1 and SCT-2
    with CPS-2 are very close to the upper bound ones with ``oracle
    selection", even both the training and development mixtures
    of target domain are taken as unlabeled ones.
    Such promising results indicate the effectiveness of our
    proposed SCT for improving the unsupervised cross-domain speech
    separation.

    \item [2)] Model robustness: Under all SCT cases, the absolute performance
    gains achieved by the adapted Conv-TasNet are much bigger than the ones
    from the adapted DPCCN. However, the best DPCCN is always better than
    the best Conv-TasNet, this is possibly due to the better robustness or
    generalization ability of our previously proposed DPCCN in \cite{dpccn}.

    \item [3)] Pseudo label selection criterion: The CPS-2 performance
        is better than CPS-1 in almost all conditions, which tells us that
        introducing \verb"mSCM" constraint is helpful to alleviate the
        pseudo ground-truth errors that brought by CPS-1. Combing both
        \verb"SCM" and \verb"mSCM" in CPS-2 can produce better high confidence pseudo labels.

    \item [4)] Cross-knowledge adaptation: Together with CPS-2, the SCT-2 achieves better
        results over SCT-1, either for the best Conv-TasNet results or for
        the DPCCN ones. It proves the importance of cross-knowledge
        adaptation for leveraging the complementary information between
        heterogeneous models to  target domain models.

    \item [5)] Number of SCT iteration: For SCT-1, both Conv-TasNet and DPCCN
       are continuously improved in the first two iterations. For SCT-2 and
       SCT-3,  Conv-TasNet still benefits from the 2nd iteration, while
       DPCCN only needs one iteration model adaptation to achieve the best results. This phenomenon indicates that the
       complementary cross-knowledge between different models can help DPCCN converge faster and achieve better performance.

     \item [6)] Necessity of two-stage CPS-2: With CPS-2, SCT-3 doesn't bring any
         improvements over SCT-2, it means that the 2nd CPS-2 stage in
         SCT-3 is useless. Possibly because the updated Conv-TasNet has
         been refined by the first stage CPS-2 outputs, the new individual separation hypothesis of this updated model has homogeneous
         acoustic characteristic with the ones in the first stage CPS-2,
         resulting in relatively simple and partial separated pseudo ground-truth in the 2nd stage CPS-2.
         Considering this phenomenon, we stop trying more CPS stages and iterations in SCT pipelines, as feeding more homogeneous data is time-consuming and hard to bring additional benefits.   
\end{enumerate}

\subsection{Performance Evaluation of SCT on WHAMR!}
\label{subsec:esw}

\begin{table}[!htbp]
  \caption{ SDRi/SI-SNRi(dB) performance on WHAMR! test set with SCT-2.}
  \label{tab:sct_whamr}
  \renewcommand\arraystretch{1.25}
  \centering
 \begin{threeparttable}
  \setlength{\tabcolsep}{0.9mm}
  \begin{tabular}{l|c|c|c|c}
    \hline
	SCT & System & \#Iter & CPS-2 & Oracle\\
	\hline
	\multirow{4}{*}{SCT-2} & \multirow{2}{*}{Conv-TasNet} & 1 & 8.28 / 7.85  & 8.64 / 8.28 \\
	& & 2 &  8.48 / 8.06 & 8.68 / 8.27 \\
	\cline{2-5}
	& \multirow{2}{*}{DPCCN} & 1 &  9.26 / 8.81 & 9.31 / 8.86 \\
	& & 2 &  8.84 / 8.40 & 8.95 / 8.52 \\
  	\hline
  \end{tabular}
  \begin{tablenotes}
  \footnotesize
  \item ``Oracle'' and $\eta$ have the same meaning as in Table \ref{tab:sct_aishell}.
  All source models are well pre-trained on Libri2Mix.
  The best setup of $\{\alpha,\beta,\eta\}$ are $\{8,5,8\}$ and $\{12,5,8\}$ in the 1st and 2nd
  SCT iteration, respectively.
  \end{tablenotes}
\end{threeparttable}
\end{table}

As the SCT-2 with CPS-2 achieves the best results in Table \ref{tab:sct_aishell},
we generalize this framework to Libri2Mix-WHAMR! (source-target) task for
a further investigation of unsupervised cross-domain speech separation.
Both source and target domain are
English speech mixtures but with different acoustic environments.
Results are shown in Table \ref{tab:sct_whamr}. It's clear that we can obtain consistent observations from this table with the ones on Aishell2Mix, which
verifies the good robustness and generalization ability of
SCT under different cross-domain speech
separation tasks. This nature of SCT is very important
for real unsupervised speech separation applications.
Our following experiments and analysis are all based on
the best SCT variant, SCT-2 with CPS-2, unless otherwise stated.

\subsection{Overall Performance Evaluation}

\begin{table}[!htbp]
  \caption{ Overall SDRi/SI-SNRi(dB) performance with different configurations.}
  \label{tab:sum_all}
  \renewcommand\arraystretch{1.25}
  \centering
 \begin{threeparttable}
  \setlength{\tabcolsep}{0.4mm}
  \begin{tabular}{l|c|c|c|c}
    \hline
	Dataset & System & Baseline & SCT & Supervised\\
	\hline
	\multirow{2}{*}{Aishell2Mix} & Conv-TasNet & 2.57/2.08 & 6.15/5.52  & 9.00/8.32 \\
	& DPCCN & 5.78/5.09 &  6.48/5.82 & 8.86/8.14 \\
  	\hline
	\multirow{2}{*}{WHAMR!} & Conv-TasNet & 6.83/6.45 & 8.48/8.06  & 11.03/10.59 \\
	& DPCCN & 8.99/8.50 &  9.26/8.81 & 11.01/10.56 \\  
  	\hline	
  \end{tabular}
  \begin{tablenotes}
  \footnotesize
  \item ``Baseline'' means model trained on source domain Libri2Mix while evaluated on target domain Aishell2Mix and WHAMR!.  
  ``SCT'' is the best adaptation configuration, i.e. SCT-2 with CPS-2.
  ``Supervised'' means model trained with ground-truth labels.
  \end{tablenotes}
\end{threeparttable}
\end{table}

To better understand the proposed SCT,  we re-organize the key experimental results in Table \ref{tab:sum_all} for an overall comparison, including results of cross-domain baselines (in Table \ref{tab:cross_domain}), the best SCT configuration (SCT-2 with CPS-2, in Table \ref{tab:sct_aishell} and \ref{tab:sct_whamr}), and the supervised results (upper bound) that trained with ground-truth labels (in Table \ref{tab:supervised}).
It's clear that the proposed SCT improves cross-domain separation performance significantly.
Compared with Conv-TasNet, the SCT gain of DPCCN is much smaller. This may because the baseline performance of Conv-TasNet is much worse, when adapted with pseudo-labeled data pairs,
Conv-TasNet will gain much more benefits.
Besides, either for Conv-TasNet or DPCCN, the selected data during SCT actually has similar acoustic characteristics. This means that after SCT adaptation, the target domain performance of Conv-TasNet and DPCCN would reach to a similar level (as shown in the SCT column).
In addition, results in Table \ref{tab:sum_all} indicate that there is still a big performance gap between SCT and the upper bound ones, which motivates us to further improve the current SCT in our future works. 
Even though, considering the huge performance gain of SCT over baseline, 
we still believe the SCT is promising for tackling unsupervised speech separation tasks.

\subsection{Heterogeneous Separation Results Fusion}
\label{subsec:srf}

\begin{table}[!htbp]
  \caption{Performance of heterogeneous separation results fusion on Aishell2Mix and WHAMR! test sets.}
  \label{tab:fusion}
  \centering
  \begin{tabular}{l|c|c|c}
    \hline
    Dataset & $\lambda$ & SDRi (dB) & SI-SNRi (dB) \\
  	\hline
  \multirow{3}{*}{Aishell2Mix} & 0.5 & 6.55 & 5.93 \\
    & 0.8 & \bf{6.57} & \bf{5.93} \\
    & 0.9 & 6.53 & 5.88 \\
    \hline
  \multirow{3}{*}{WHAMR!} & 0.5 & 9.19 & 8.79 \\
    & 0.8 & \bf{9.34} & \bf{8.92} \\
    & 0.9 & 9.32 & 8.89\\
    \hline
  \end{tabular}
  %\vspace{-0.3cm}
\end{table}

Motivated by the design of SCT, we believe that the separation results of
the final adapted target domain
models also have complementary information,  because
they are derived from two different neural networks with heterogeneous
structure. Therefore, a simple linear fusion of separated signal spectrograms
is preliminarily   investigated to further improve the SCT.

Results are shown in Table \ref{tab:fusion}, where $\lambda$
and $1-\lambda$ are linear weights for the signal spectrograms of
adapted DPCCN and Conv-TasNet outputs respectively. The pairwise cosine similarity
is used to find the best match spectrograms that belong to
the same speaker during linear fusion.
Compared with the best SCT-2 results in
Table \ref{tab:sct_aishell} and \ref{tab:sct_whamr}, this simple fusion is still
able to bring slight performance improvements. This indicates that, it
is possible to exploit the complementary information between
SCT outputs to further improve the final separation results.
It will be interesting to try other and more effective separation results
fusion methods in future works.

\subsection{Data Quantity Analysis of Pseudo Ground-truth}
\label{subsec:sqa}

\begin{figure}[!htbp]
  \centering
  \includegraphics[width=8cm]{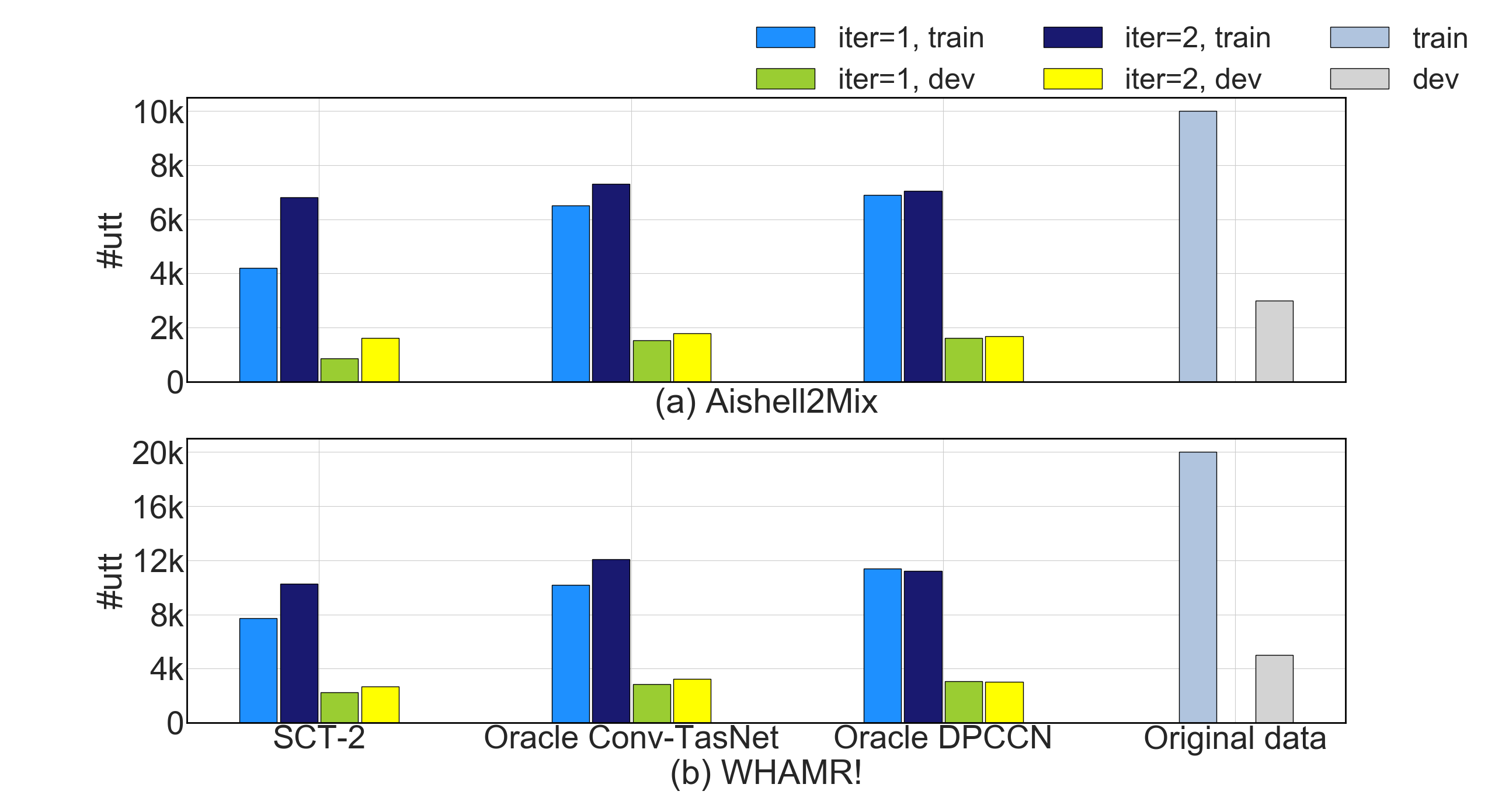}
  \caption{Data quantity of selected pseudo ground-truth of
            SCT-2 (with CPS-2) versus the ``Oracle selection" on Aishell2Mix and WHAMR!
            unlabeled training and development sets. }
  \label{fig:spk_bar}
  %\vspace{-0.5cm}
\end{figure}

The key success of the proposed SCT depends on the high
confidence pseudo-labeling. It's very important to analyze the
data amount statistics of the selected pseudo ground-truth during SCT in
different unsupervised separation tasks.
Fig.\ref{fig:spk_bar} shows the statistics that 
used to adapt the heterogeneous networks during each iteration of
SCT-2 (with CPS-2) in Table \ref{tab:sct_aishell} and \ref{tab:sct_whamr},
including the selected training and development
data of unlabeled Aishell2Mix and WHAMR! datasets.
For further comparisons, we also show the corresponding upper bound
data statistics generated by the ``Oracle selection" as references.
Note that, as the cross-knowledge adaptation is applied
during SCT-2, the data amounts of ``D-Pseudo Labeled Set"
and ``T-Pseudo Labeled Set" are the same but with different
ground-truth individual signals, so we use ``SCT-2" to represent
both of them, and
the ``Oracle Conv-TasNet'' and ``Oracle DPCCN'' in
Fig.\ref{fig:spk_bar} actually represent the oracle
amount of pseudo data that selected to adapt
the Conv-TasNet and DPCCN, respectively.

From Fig. \ref{fig:spk_bar}, three findings are observed:
1) the 2nd SCT-2 iteration can produce more high confidence data, and
the selected data quantity is close to the upper bounds with ``Oracle selection", indicating the heterogeneous structure in SCT and
the thresholds of CPS-2 are reasonable;
2) on Aishell2Mix, both the selected training and development
data increments in the 2nd  iteration are higher than the ones
on WHAMR!, which means the multiple SCT-2 iterations are
necessary for tasks with the larger cross-domain mismatch.
3) for ``Oracle DPCCN'', the selected data quantities of two iterations are almost the same, indicating the pseudo-labeled
mixtures in each iteration are a large number of homogeneous data that
results in an over-trained DPCCN model. This is also the reason of worse results in the 2nd iteration that shown 
in Table \ref{tab:sct_aishell} and \ref{tab:sct_whamr}.
All these above findings give a well support to the separation results as presented in both Table \ref{tab:sct_aishell} and \ref{tab:sct_whamr}.

\subsection{Gender Preference Analysis}
\label{subsec:gpa}

\begin{figure}[!htbp]
  \centering
  \includegraphics[width=8cm]{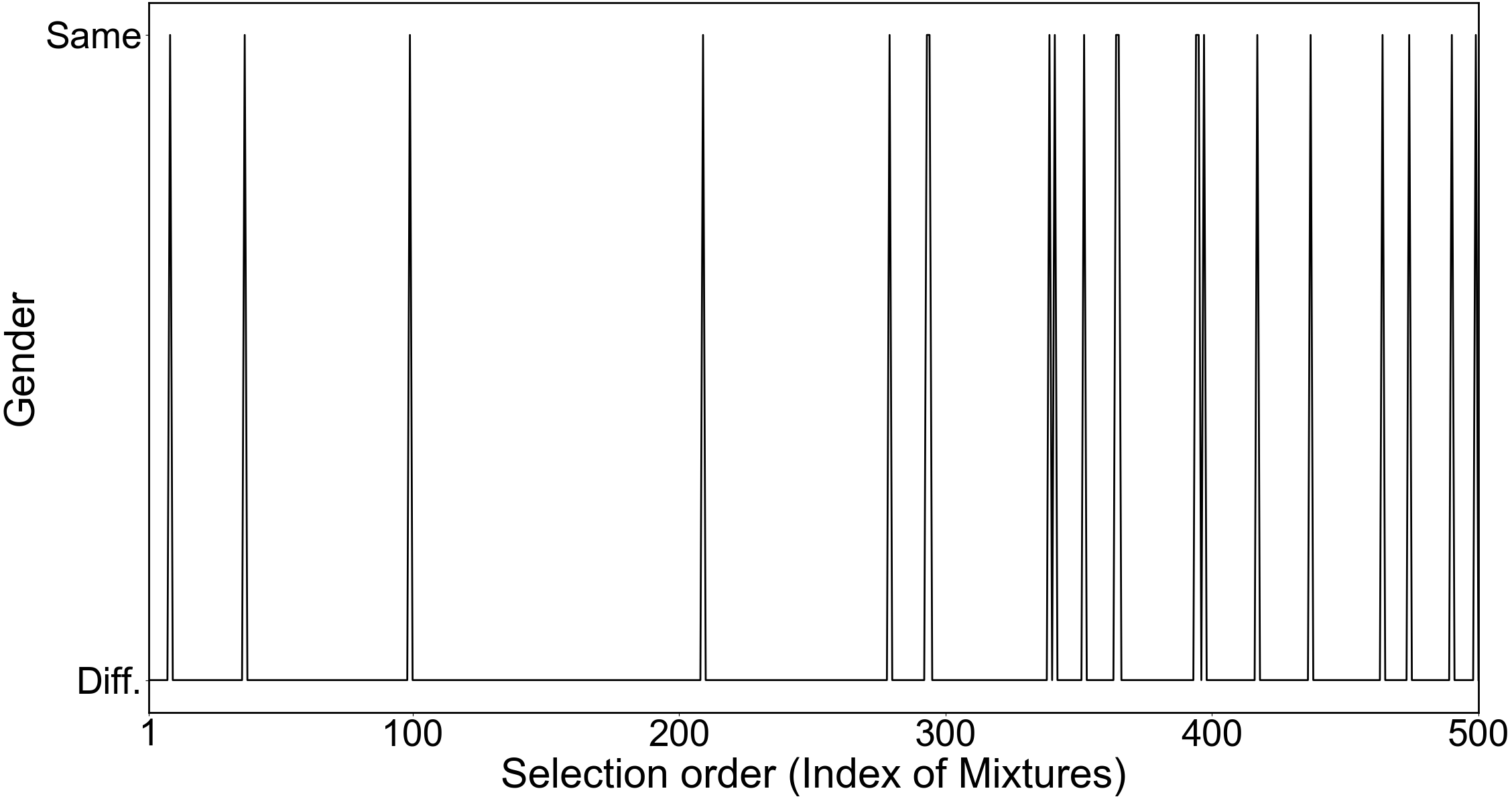}
  \caption{Gender information of top 500 CPS-2 results on
  Aishell2Mix development set. ``Diff.'' represents the different gender.}
  \label{fig:gender}
  %\vspace{-0.5cm}
\end{figure}

As we all know, the speech mixed with different gender speakers
is easier to separate than that with the same gender speakers.
In this section, we investigate the gender distribution of selected
pseudo-labels on the Aishell2Mix development set. The gender information of top 500  mixtures with the best CPS-2 setup, $\alpha=8$ and $\beta=5$,
is presented in Fig. \ref{fig:gender}, where each spike pulse
represents the gender in each mixture changing from different to the
same.

From Fig.\ref{fig:gender}, it's clear that the proposed CPS-2
prefers to select the mixtures with different gender speakers.
The sparse spike pulse shows the extremely low proportion of
same gender mixtures in the entire selected speech, and
its distribution tends to denser when the confidence of the 
selected mixture becomes lower (larger selection order).
These phenomena are consistent with our prior knowledge,
i.e., the speech mixed by different gender speakers is easier to separate
and its separated signals from heterogeneous models
show a higher separation consistency.

\subsection{Bad Cases Analysis}
\label{subsec:bca}

\begin{figure}[!htbp]
  \centering
  \includegraphics[width=8cm]{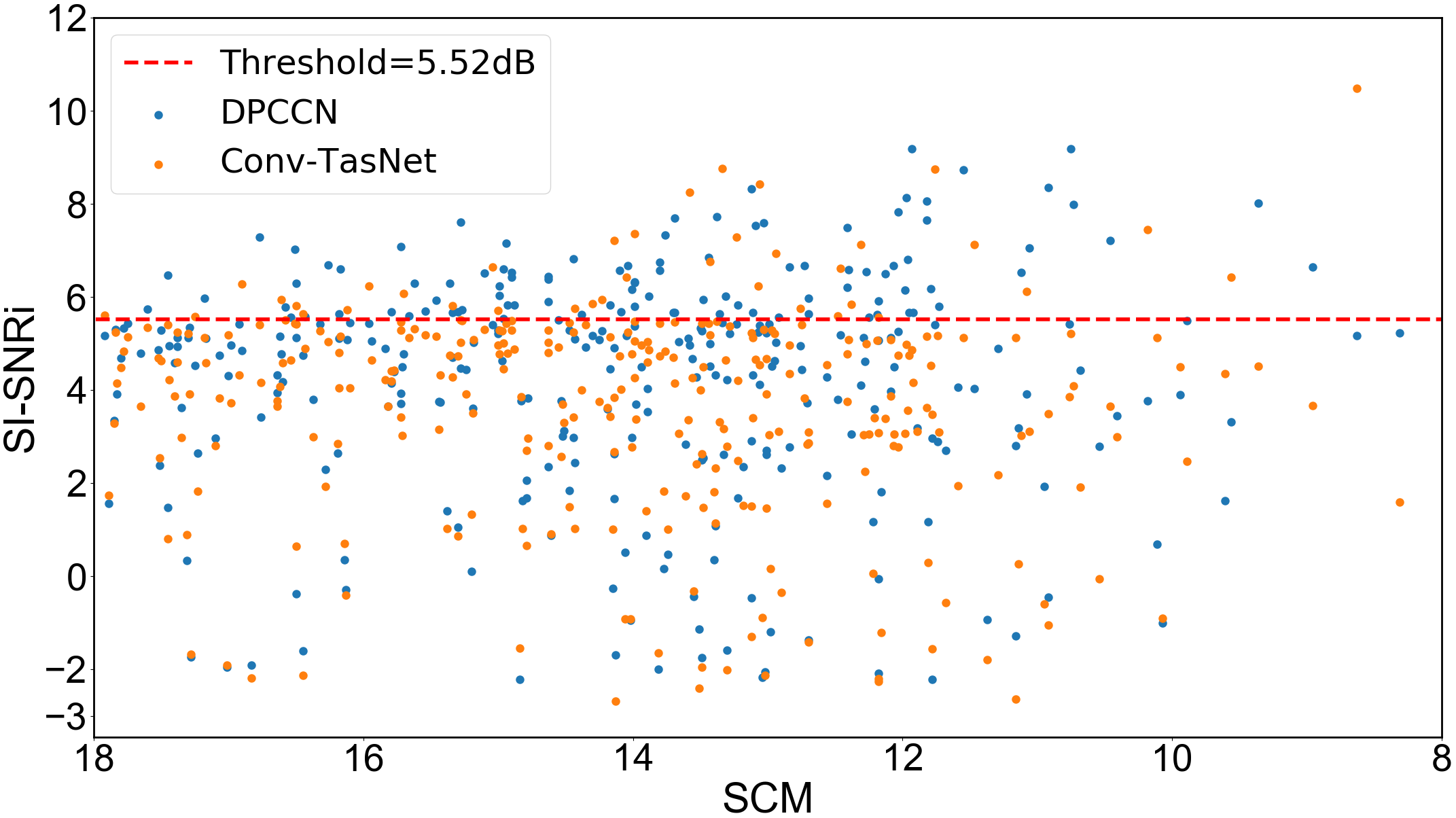}
  %\caption{Histogram of bad cases in SCT-2 (with SCS-2) results.}
  \caption{SI-SNRi (dB) of DPCCN and Conv-TasNet separation results
   of the 310 ``bad cases" varies with the separation consistency measure.}
  \label{fig:bad_hist}
\end{figure}

Finally, we perform a bad cases analysis of the separation results on the Aishell2Mix development set in Fig.\ref{fig:bad_hist}. All these
unlabeled mixtures in this dataset are first separated by
the best adapted target domain DPCCN and Conv-TasNet
models in Table \ref{tab:sct_aishell} (SCT-2 with CPS-2). Then
the CPS-2 with $\alpha=8$, $\beta=5$ is used to select the
pseudo labels and 1716 mixtures' \verb"SCI" tuples are selected in total.
Next, we calculate the standard separation performance (SI-SNRi) of both
the DPCCN and Conv-TasNet separation outputs by taking the real ground-truth to evaluate each mixture performance, and we refer them to $\mathrm{\operatorname{SI-SNRi}_{DPCCN}}$ and
$\mathrm{\operatorname{SI-SNRi_{Conv-TasNet}}}$ for simplicity. Then, we compare each SI-SNRi
with the average SI-SNRi (5.52 dB, the best performance of Conv-TasNet
in Table \ref{tab:sct_aishell}) of Aishell2Mix test set to determine
whether the current mixture separation is a ``bad case" or not.
For each selected mixture, if its  $\{\mathrm{\operatorname{SI-SNRi_{DPCCN}~||~  SI-SNRi_{Conv-TasNet}}} \} < 5.52~\mathrm{dB}$, we consider it a failure separation (F) and
the corresponding mixed speech is taken as a ``bad case", otherwise we take it as a succuss separation (T).
With this rule, total 310 of 1716 ($18.1\%$) mixtures are taken as ``bad cases".

The reason behind this ``bad case" decision rule is that,
in the speech separation field, there is no measurement to evaluate each
speech separation is 100\% accurate or not. Therefore, we think that,
the real separation performance of the best separation model can be taken as a proper heuristic signal distortion threshold
for a rough ``bad case" analysis. And in our SCT-2, when compared with
the best DPCCN performance (5.82 dB) in Table \ref{tab:sct_aishell},
the Conv-TasNet performance, 5.52 dB is a stricter one for the
``bad case" decision.

Fig. \ref{fig:bad_hist} shows how the DPCCN and Conv-TasNet
separation outputs of the 310 ``bad cases" SI-SNRi
varies with the separation consistency \verb"SCM". From these scatter points,
we see that, with our proposed CPS-2, the selected 310 mixture
pseudo labels still contain low-quality ones that are not suitable to be taken
as ground-truth, even though all these mixtures have relatively high consistency
confidence. From the left part of this figure,
we find some ``bad cases" with high separation
consistency $\verb"SCM" > \mathrm{12~dB} $ but their real separation performances
are very low (SI-SNRi $< \mathrm{2~dB}$). However, on the contrary,
the right part figure shows some low \verb"SCM" mixtures are also separated
very well. Therefore, we 
speculate that, these
``bad cases" may not be too bad if they are within the error tolerance of system training data, they may be taken as small
noisy distortions of the whole pseudo labeled
training set and may help to enhance the model robustness.
That's why we still obtain promising performance in Table \ref{tab:sct_aishell}
using the proposed SCT.

\begin{figure}[!htbp]
  \centering
  \includegraphics[width=8cm]{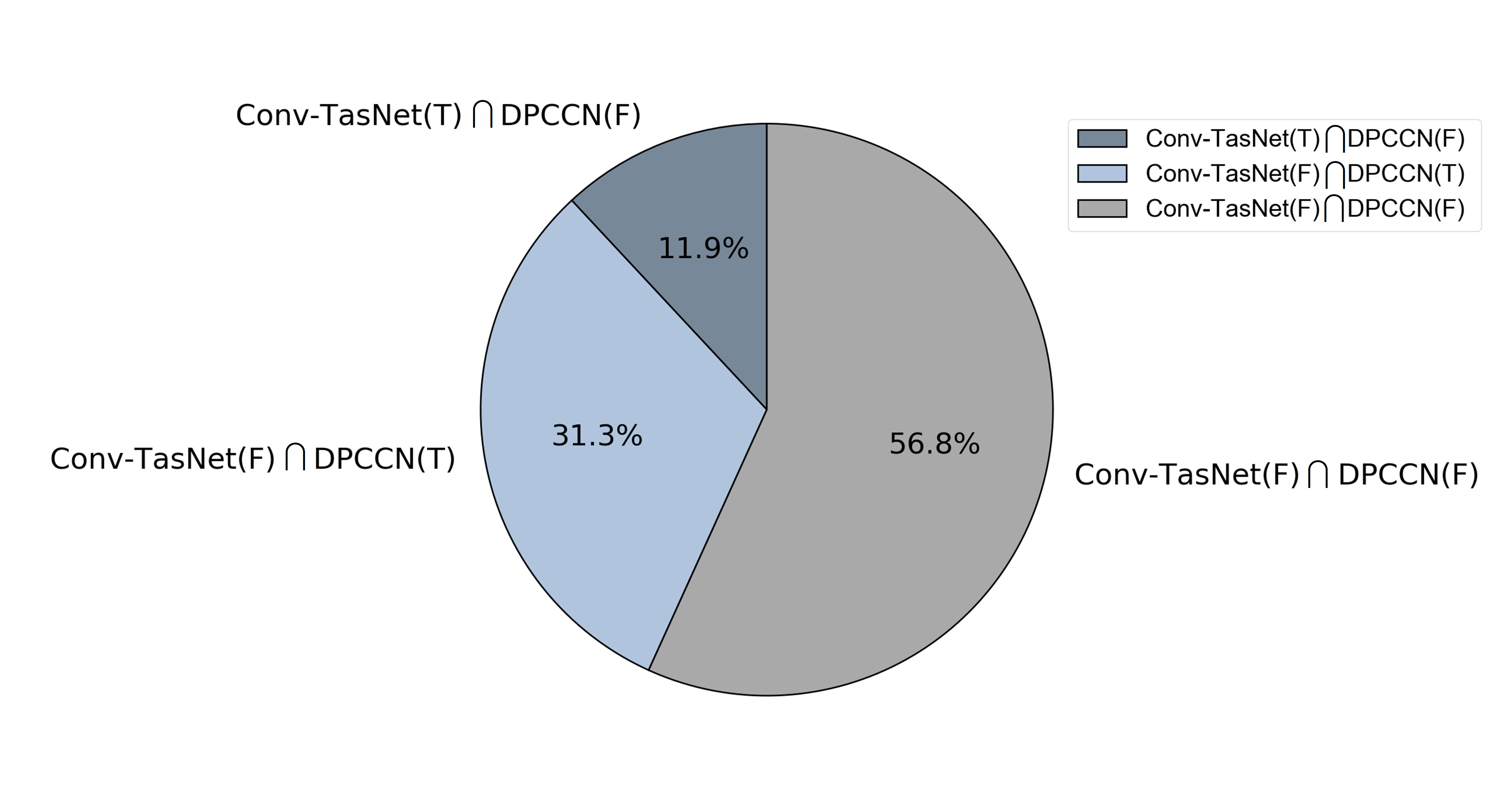}
  \caption{Detailed separation statistics of the 310 ``bad cases" in
  Aishell2Mix development set.}
  \label{fig:bad_pie}
\end{figure}

Fig.\ref{fig:bad_pie} demonstrates other detailed separation statistics
of the same 310 ``bad cases" on Aishell2Mix development set from another
perspective. The \verb"T,F" means the success, failure separation
as defined in the above statements. Each ``bad case" covers three kinds of
\verb"T,F" combination, such as, $\verb"Conv-TasNet(T)" \cap \verb"DPCCN(F)"$ means for each
unlabeled mixture, the separation of Conv-TasNet is success while DPCCN
is failure.

From Fig.\ref{fig:bad_pie}, we see  56.8\% of these ``bad cases" are
consistent failure separations for both DPCCN and Conv-TasNet.
However, there is still around half of the data can be separated well
by one of these two heterogeneous systems, as shown in the two
\verb"T" $\cap$ \verb"F" combinations. This observation clearly proves the large
complementary information between two heterogeneous separation models, as
the time-domain Conv-TasNet and the time-frequency domain DPCCN used in
our SCT. And it also inspires us to improve the SCT-1 to SCT-2 using the
cross-knowledge adaptation.
Besides, for the 31.3\% vs 11.9\% \verb"T" $\cap$ \verb"F" combination,
we see there are much more DPCCN success mixture separations than
the Conv-TasNet on this difficult-to-separate 310 mixtures.
This means DPCCN is a better candidate for robust speech separation task,
using DPCCN as the primary model and its outputs as references in the whole SCT process
is reasonable.

\section{Conclusion}
\label{sec:conclusion}

In this paper, we proposed an iterative separation consistency training (SCT)
framework, a practical source model adaptation technology for
cross-domain unsupervised speech separation tasks.
By introducing an effective pseudo-labeling approach,
the unlabeled target domain mixtures are well exploited
for target model adaptation, which successfully reduces
the strong ground-truth reliance of most state-of-the-art
supervised speech separation systems.

Different from previous works, SCT follows a heterogeneous structure,
it is composed of a masking-based  time-domain separation model, Conv-TasNet,
and a mapping-based  time-frequency domain separation model, DPCCN.
Due to this heterogeneous structure and the specially designed
separation consistency measures, SCT can not only
perform the pseudo-labeling of unlabeled mixtures
automatically, but also can ensure the selected pseudo ground-truths
are high quality and informative.
Moreover, by introducing
the cross-knowledge adaptation in SCT, the 
large complementary information between heterogeneous models
is maximally leveraged to improve the primary separation system.
In addition, the iterative adaptation nature in SCT provides an increased
chance to improve the primary model
when there is a large amount of unlabeled mixtures available.
Finally, we find this heterogeneous
design of SCT also has the potential to further improve the
final separation system performance by combing two final adapted separation
model at the level of their outputs.

We verified the effectiveness of our  proposed methods on two cross-domain conditions:
the reverberant English and the anechoic Mandarin acoustic environments.
Results show that, under each condition, both the heterogeneous separation models
are significantly improved, their target domain performance is very close to the
upper bound ones, even the target domain training and
development sets are all unlabeled mixtures.
In addition, through the bad case analysis, we find that
the SCT will definitely introduce some error pseudo ground-truth
to a certain extent. 
This limitation of current SCT still needs to be 
improved in our future works before we apply it to real 
speech separation applications. 

\section*{Acknowledgment}

This work is supported by the National Natural Science Foundation of China (Grant No.62071302 and No.61701306).

\bibliographystyle{elsarticle-num}
\bibliography{mybib}

\end{document}